 \documentclass{elsart}          
 \usepackage{graphicx}      

\begin{document}

 \newcommand{\JPA}{J. Phys. A} 
 \newcommand{\JPC}{J. Phys. C} 
 \newcommand{\JSP}{J. Stat. Phys.}   
 \newcommand{\PA}{Physica A}   
 \newcommand{\PRB}{Phys. Rev. B}  
 \newcommand{\PRE}{Phys. Rev. E}  
 \newcommand{\PRL}{Phys. Rev. Lett.}   

 \newcommand{\be}{\begin{equation}}
 \newcommand{\bea}{\begin{eqnarray}}
 \newcommand{\cF}{{\mathcal F}}
 \newcommand{\cG}{{\mathcal G}}
 \newcommand{\cH}{{\mathcal H}}
 \newcommand{\cO}{{\mathcal O}}
 \newcommand{\cq}{complex-$q$ }
 \newcommand{\cS}{complex-$S$ }
 \newcommand{\ee}{\end{equation}}
 \newcommand{\eea}{\end{eqnarray}}
 \newcommand{\etl}{{\it et al.}\,\,}   
 \newcommand{\req}{real-$q$ }
 \newcommand{\vfi}{\varphi}

   \renewcommand              
   \baselinestretch 1              
   \baselineskip 15pt              %
  \setlength\textwidth{19cm}      
  \setlength\textheight{23cm}     

 \begin{frontmatter} 

 \title{Complex-$q$ zeros of the partition function of the 
        Potts model with long-range interactions}  

 \author{Zvonko Glumac\thanksref{corresp}}, 
 \thanks[corresp]{Corresponding author. tel.: (385)-031-208-900;
                                        fax:  (385)-031-208-905; 
                                        e-mail: zvonko@vrabac.ifs.hr.} 

 \address{Faculty of Electrical Engineering, 
         Kneza Trpimira 2B, 31 000 Osijek, Croatia}  
 \author{Katarina Uzelac}

 \address{ Institute of Physics, P.O.B. 304,
 Bijeni\v{c}ka 46, HR-10000 Zagreb, Croatia}

 \begin{abstract}
The zeros of the partition function of the ferromagnetic
$q$-state Potts model with long-range interactions in the \cq plane
are studied in the mean-field case, while preliminary numerical results are 
reported for the finite $1d$ chains with power-law decaying interactions. 
 In both cases, at any fixed temperature, the zeros lie on the 
arc-shaped contours, which cross the positive real axis at the
value for which the given temperature is transition temperature. 
For finite number of spins the positive real axis is free of zeros, which 
approach to it in the thermodynamic limit.
 The convergence exponent of the zero closest to the positive \req axis is 
 found to have the same value as the temperature critical exponent $1/\nu$.

 \leftline{PACS: 05.50.+q, 64.60.Cn} 
 \end{abstract}


\end{frontmatter}

  \section{Introduction}  \label{sec-int}  

 Several decades ago, it was shown that the study of distribution 
 of the partition-function zeros in the complex plane of an appropriate 
variable,  can provide relevant  informations on phase transitions.
 In their pioneering work, Lee and Yang \cite{YL52,LY52} have shown 
 the connection between the existence of phase transition and the 
 distribution of partition function zeros of the ferromagnetic Ising model 
 in the complex plane of a symmetry breaking field.
 Later on, \cite{F65,OKSK68},  similar connection was established for 
 the partition function zeros in the complex-temperature plane. 
In both cases, the singularities at the transition point 
 are closely related to the way that the zeros approach (either with 
temperature or with the size of the system) to the positive real axis.

The partition function zeros were investigated for number of models, 
including the Potts model both in complex-field \cite{KC98,KC99} and 
in complex-temperature  \cite{B90,M91,CHW96,KCCH99} plane.
In the Potts model a relevant quantity is also the number of states 
$q$, often considered as a continuous parameter, which is formally 
possible on the basis of its graphical representation \cite{FK69} as it 
will be discussed later in text.  
This gives the motivation to investigate the zeros for this model also 
in the plane of complex $q$.   
Such an idea was already formulated \cite{Bx86,Bx87} but mostly in 
context of chromatic polynomials in connection with studies of the 
ground state of the anti-ferromagnetic Potts model 
 \cite{ST97a,ST97b,RST98,ST99a,ST99b,S99,SS00}.
Only a few works on complex-$q$ plane address the ferromagnetic 
Potts model at finite temperatures.
The papers by Chang and  Shrock, \cite{CS00}, study the loci of 
partition function zeros in $2d$ Potts model with different boundary 
conditions.
In their very recent paper Kim and Creswick \cite{KC01} give a systematic
study of both Fisher and complex-$q$ zeros in the $2d$ Potts model with
short-range  interactions and find the similarity between the scaling property
of complex-$q$ zeros and the den Nijs expression for the thermal critical
exponent.

  In the present paper we investigate the zeros in the complex-$q$ plane
  in connection with the first- and second-order phase
  transitions in the ferromagnetic Potts model.
  To this purpose we consider here the Potts model with long-range 
power-law decaying interactions, which
already in one dimension displays diversity of critical regimes
and embraces most of the nontrivial aspects of phase
transitions, characteristic of the Potts model in general.
The limiting case of this model, where all the interactions are equal
(and which represents the mean-field case of the model) has the
advantage to be exactly solvable for arbitrary $q$.

In the next section we describe briefly the starting
 Hamiltonian and its graphical representation. In subsequent sections
 we present the results for the mean-field (Section 3) and the
 power-law decaying
 interactions case (Section 4).

\section{Hamiltonian}

The ferromagnetic Potts model with interactions of arbitrary range is 
described by the Hamiltonian
 \be \label{eq:lrham}
  H  = - \sum_{i,j}\; J_{i,j}\cdot
             \delta\,(s_i, s_{j}),   
 \ee  
where $s_i$ denotes the Potts spin at site $i$ of a 
 $d$-dimensional lattice, 
which can take $q$ values, and $J_{i,j}\ge 0$ denotes the ferromagnetic 
interaction between spins at sites $i$ and $j$.

An impressive amount of studies of this model in the case of 
nearest neighbour interactions (for a review see Wu \cite{W82})
reveal its very complex critical behaviour which varies with $q$ and
comprises  transitions of different orders and belonging to various 
universality classes. 

The model with long-range interactions was studied less, 
and mostly in two special cases.
The first is the case when all the interactions are equal and 
it represents the mean-field limit of the model. 
It can be approached by the saddle point approximation 
 \cite{KMS54,SF73,MS74,PL76} and in the thermodynamic limit 
the exact expressions, in particular those for critical 
temperature and exponents may be derived for all values of $q$.
Second is the case with power-law interactions 
decaying with distance as $1/r^{d+\sigma}$. 
The type of phase transition there depends in addition 
on the parameter of range  $\sigma$, which in certain aspects has 
similar effects as the change of dimensionality \cite{LB96}. 
The phase transition is there present in one dimension
 \cite{D69,ACCN88,GU93} and by
varying $\sigma$, series of different regimes are met, which 
makes this $1d$ model a useful paradigm for most of important 
features of the model in general (including also the first-order 
phase transition).  
It was a subject of  number of recent studies
\cite{GU93,LB95,C95,UG97,GU98,M99,BDD99,KL00,LM01}. 

	The graphical representation of the Potts model \cite{FK69} 
holds for arbitrary range of interactions.
The partition function for the Hamiltonian (\ref{eq:lrham}) 
may be expressed as 
  \be \label{eq:zlropci}
   Z_N = \sum_{{\rm all}\;\cG} \,
   \prod_{\rm {active\atop links}} v_{i,j} \;\,q^{n(\cG)},
   \ee
where the summation is taken over all possible graphs $\cG$.
Each graph represents one possible configuration of 
connections between the $N$ spins, called active links, 
each of which contributes with a factor $v_{i,j} = \exp(J_{i,j}/{k_B\,T}) - 1$. 
$n(\cG)$ denotes the number of  
of disconnected parts of the graph called clusters.
The product is taken over all the active links of a considered graph.

Whatever the interactions are, the number of clusters that a single graph 
contains ranges from one cluster  (low temperature limit, where 
all spins are interconnected), up to at most $N$ clusters
(high temperature limit where no active  links are present and each 
spin is a cluster of size one).
Thus, by collecting all the graphs with the same number of clusters, 
the partition function becomes simply an $N$-th order polynomial in $q$
   \be \label{eq:Npol}
   Z_N = \sum_{n = 1}^N\,a_n(T,J_{i,j},N)\;q^n.
   \ee
This made possible to treat the number of states $q$ as a continuous parameter.
It also gives the possibility to examine the zeros of the partition function
as the zeros of the $N$-th order polynomial in $q$.  

The coefficients $a_{n}$ are functions of temperature, the finite size 
and interactions, and are quite complicated in the general case 
(\ref{eq:lrham}), but can be written in more compact form for  
the special cases considered here.

\section {The mean-field case}

The simplest case of long-range interactions is the one where all the 
interactions are set to be equal, $J_{i,j}=J$, which reduces the  
Hamiltonian (\ref{eq:lrham})  to the form
 \be \label{eq:mfham}
 \cH_{MF} = H_{MF}/{k_B\,T} = -
  \frac{K}{N} \sum_{i=1}^{N-1}\;\sum_{j=1}^{N-i}\; 
             \delta\,(s_i, s_{i+j}),   
 \ee  
where $K=J/{k_B\,T}$, and division by $N$ is necessary to keep 
the summation finite. 
This model corresponds to the mean-field limit. It was solved by the 
saddle point approximation \cite{KMS54}. 
The phase transition is there of the first order for $q > 2$ 
and of the second order for $q \le 2$.
The exact expressions for the inverse of the transition temperature 
$K_t$  is given by 
 \be \label{eq:Kexact1}  
  K_t =  2\,\frac{q-1}{q-2}\,\ln (q-1), 
\hspace{1cm} 
 \ee 
for $q > 2$, 
and by $K_c = q$ for $q \le 2$.   
Exact analytical expressions may be derived also for the 
critical exponents matching the values obtained in the
mean-field regime of the short-range interaction model \cite{PL76}
(except those related to the correlation function, as it 
will be discussed later in text). 

The graph representation (\ref{eq:zlropci}) reduces in the mean-field limit to
   \be \label{eq:zmf}   
   Z_N = \sum_{{\rm all}\;\cG} \, v^{b(\cG)}\,q^{n(\cG)},
   \ee
where the active links have all the same strength $v = \exp(K/N) - 1$, 
  independently of distance between the spins they connect and
 $b(\cG)$ denotes the number of active links in a given graph $\cG$. 
Written as a  polynomial in $q$  the above equation is equal to
   \be \label{eq:Npolmf}
   Z_N = \sum_{n = 1}^N\,a_n(v, N)\;q^n,
   \ee
 where the coefficients $a_n$ are real and positive and depend on 
temperature and size $N$ only. 
(Notice, that, since the number of active bonds may reach up to $N\,(N-1)/2$, 
the above partition function could equally be written as a polynomial 
in temperature variable $v$, which would be of order $N\,(N-1)/2$.)

\subsection{Numerical results}

For relatively small sizes the evaluation of coefficients and 
finding the zeros of the polynomial (\ref{eq:Npolmf}) in the 
complex plane may easily be performed numerically.
Numerical results for sizes up to $N = 35$ spins are presented in
 Figs. 1 (a)-(c) for the three different temperatures,  
  $K = 0.5, 2, 4.540457$.   
These temperatures are the transition temperatures of the mean-field model
with $q = 0.5, 2$, and $8$ respectively.
Each figure displays the data obtained for several sizes.

One may observe that in all the three cases the zeros lie on contours
similar but different from circle
which cross the real axis near the value of $q$, for which the given 
temperature is the transition temperature. 
The zeros are absent from the positive real axis, but approach to 
it with increasing $N$.

We were thus interested to perform a more precise extrapolation of this
crossing point in the thermodynamic limit and studied the convergence
of the zero closest to the positive real axis 
with $N$.
Further, we have examined the possible connection of the convergence 
of the closest zero with $N$ to the scaling properties of the model 
near criticality.

To this purpose we have used Burlisch and Stoer (BST) extrapolation
procedure \cite{BS64,HS88}.
This algorithm was shown \cite{HS88} to converge rapidly, needs a modest 
amount  of data and, in  advantage over the similar algorithm  by van den 
Broeck and Schwartz (VBS) \cite{VBS79}, it is less  sensitive to the rounding 
errors.

   In short, BST algorithm extrapolates a sequence of the form
   \be \label{eq:seq}  
   A(N) = A_{\infty} + a_1\,\cdot\,N^{-\omega} + 
   a_2\,\cdot\,N^{-2\,\omega} 
   + \cdots  
   \ee 
   in the $N\,\to\,\infty$ limit by a set of transformations recursively 
   given by 
   \bea 
   A_m^{(N)} & = & A_{m - 1}^{(N + 1)}  \\
   & + & \left(A_{m - 1}^{(N + 1)}  - A_{m - 1}^{(N)}\right) \, \cdot \,
   \left[ \left(\frac{N + m}{N}\right)^{\omega}\, 
   \left( 1 - 
   \frac{A_{m - 1}^{(N + 1)} - A_{m - 1}^{(N)}}{A_{m - 1}^{(N + 1)} 
   - A_{m - 2}^{(N + 1)}}
   \right) - 1 \right]^{-1},  \nonumber 
   \eea  
   where $A_{-1}^{(N)} = 0$ and $A_0^{(N)} = q_{1, 2}^{(0)}(N)$ 
   are the input data and $\omega$ is a free parameter. 

If the original series is of the form:
   \be \label{eq:seq12}  
   A(N) = A_{\infty} + c_1\,\cdot\,N^{-\omega_1} + 
   c_2\,\cdot\,N^{-\omega_2} 
   + \cdots  
   \ee 
it incorporates the leading convergence exponent as $\omega$, 
while the higher order approximants converge as $N^{-\omega^{'}}$, 
where $\omega^{'}=min(2\omega_1,\omega_2)$.
Although it was shown, that the extrapolations are insensitive to the choice 
of $\omega$ in a wide range of values, it influences its convergence
speed. Thus, by maximising the convergence of higher approximants one can in 
the same time obtain both the extrapolated value and the leading convergence 
exponent. Also, when the extrapolated value is known in advance, 
$\omega$ yields the leading convergence exponent.

We have examined the convergence of the real and imaginary parts 
of the zero closest to the real axis, denoted as $q_{1, 2}^{(0)}(N)$ 
for several  temperatures using the numerical results ranging from 
$N = 21$ to $N = 35$. 

 In the extrapolation of  both real and imaginary part
 $q_{1, 2}^{(0)}(N)$,    
 the same parameter $\omega$ was used and was fixed by the constraint 
$q_2^{(0)}(N\,\to\,\infty)\,\to\,0$.  

The results of extrapolations are presented on Table 1. 
By $Q_0$ is denoted the value of $q$ for which the considered temperature
is known to be the transition temperature. 
All the data were obtained by requiring that condition $q_2^{(0)}(N)\,\to\,0$
should be fulfilled with precision of $10^{-6}$.

   \begin{table}[hbt] \label{tb:rez12}
   \caption{ The BST extrapolation results of the positions of 
             closest zeros.}
   \begin{center} \begin{tabular}{ccccccc}
   &  &  &  &  &  & \\
   \hline
  $K$ & 0.1 & 0.5 & 1.0 & 2.0 & 2.772589 & 4.540457  \\
  $Q_0$ & 0.1 & 0.5 & 1.0 & 2.0 & 3.0 & 8.0  \\
  $q_1$ & 0.09998 & 0.504 & 0.95 & 2.05 & 3.02 & 8.00009 \\
  $\omega$ & 0.34 & 0.33 & 0.33 & 0.55 & 0.91 & 1.19   \\
   \hline
   &  &  &  &  &  & \\
   \end{tabular} \end{center} \end{table}

   The presented results indicate that: \

   (i) loci of complex-$q$ zeros intersect the positive \req axis at 
    value $q_1 = Q_0$. The values agree up to a few percents.

(ii)  Power-law convergence of the closest zero  
is observed in all the cases considered. 

The convergence exponent $\omega$ represents the scaling with 
size of the distance from the real axis to the closest zero.
Before proceeding with the finite-size scaling analysis it is important to 
notice that here the scaling is not done with the linear size, but
rather with volume, more precisely, with number of spins,
since in the infinitely coordinated model the distance as well as the
dimensionality lose their meaning.
In this respect the considered mean-field limit differs from the standard
mean-field approach to the short-range interaction model.
As pointed out by Botet \etl \cite{BJP82,BJ83} the basic scaling 
quantity in the infinitely coordinated system is not the correlation length, 
but the "correlation number"
$N_c$, which scales with temperature with the critical exponent
$\nu^{*}=\nu_{MF} \cdot d_c$. The exponent $\nu_{MF}$ and $d_c$
are, respectively, the mean-field value of the critical exponent $\nu$ 
    and 
the upper critical dimension (the dimension above which the mean-field regime
sets on) of the short-range model.

Thus, the temperature critical exponent for the present model,
where only the scaling with the number of spins can be applied
will be $1/\nu^{*}$ and not $1/\nu_{MF}$.
In the Potts model  $\nu_{MF}=1/2$ both for $q < 2$ and $q=2$, while
$d_c=4$ for $q=2$, but $d_c=6$ for $q<2$ for symmetry  reasons.
This gives  $\nu^{*}=2$ for $q=2$ and $\nu^{*}=3$ for $q<2$.

The results presented in Table 1 clearly show that
in the regime of second-order phase transition ($q \leq 2$) the 
convergence exponent coincides up to a few percents 
with the temperature critical exponent  $1/\nu^{*}$.
This result should be compared to the one in complex temperature plane, 
where it was shown \cite{IPZ83}  that the distance of the zero 
closest to the real axis scales as the temperature critical exponent.
Similar behaviour in a complex-$q$ plane was recently observed in 
the $2d$ ferromagnetic Potts model by Kim and Creswick \cite{KC01}, 
but the precision of their results did not permit them to be conclusive
whether the obtained convergence exponent was indeed equal to the 
exponent $1/\nu$. 

In cases $q=3$ and $q=8$ we obtain approximately linear convergence, 
proportional to the system size, which is the form of scaling that is 
expected for the first-order transition.

\subsection{Large $N$ expansion}

   In the limit $N \gg 1$ 
   some analytical results may easily be derived 
   by extending the mean-field solution by Kihara \etl 
   For the large number of spins $N$,
   the free energy density, without the terms which are constant 
   in the order parameter $S$, is given by  
   \bea \label{eq:slen}
    & \cF & (K, q, S)  = \\  & - &  \,S^2\,\frac{K}{2}\,\frac{q-1}{q} 
    + 
   \frac{1+(q-1)S}{q}\,\ln\,\left[1+(q-1)S\right]  +
   (q-1)\,\frac{1-S}{q}\,\ln(1-S). \nonumber  
   \eea
    The partition function is given by 
   \be 
    Z_N (K, q)  
    \sim 
    \,\int_0^1 \exp\,(- N\,\cF)\,dS,
   \ee
    where the proportionality sign stands for all the omitted constant terms.      

We limit our considerations here to the first-order regime of 
temperatures, where the free energy density at the transition 
shows two equal minima, and the partition function integral can be
   approximately solved  by a saddle point method
   \be
    Z_N \sim \, 1 + \exp\,[-\,N \cF(S_{min})\,]. \label{eq:pf}
   \ee
The first term on the right hand side denotes the minimum at $S=0$,
   and the second term is the 
contribution of the second minimum at $S=S_{min} > 0$,
   which is obtained from 
 the equation for the extremum 
   \be
   K S = \ln \frac{1+(q-1)S}{1-S}. \label{eq:ext}
   \ee

   The straightforward way to find the complex-$q$ zeros would consist
   of two steps:
    first, by using (\ref{eq:ext}) one should represent $\cF$ as a function 
of $K$ and $q$ only;
   second,  one should solve the equation $Z_N = 0$ 
by analytic continuation of $q$ to the complex plane.
   Since $\cF = \cF_R + i \cF_I$ then becomes complex, 
the vanishing of $Z_N$, as given by (\ref{eq:pf}), requires
   \be \label{eq:uvjeti}
   \cF_R   = \, 0, \hspace{1cm} N \cF_I = \, \pm \, (2 m + 1) \pi,
    \;\;\;\;m = 0, 1, \ldots
   \ee
   These two equations determine the position of complex conjugate
   pairs of zeros $(q_1^{(m)}, \pm\,q_2^{(m)})$
   in the complex-$q$ plane.

   The first of the above mentioned steps cannot be performed in the
   straightforward way.
   Instead, we will solve (\ref{eq:ext}) in variable $q$
   \be
   q = \frac{1-S}{S} [ \exp(K S) - 1 ],  \label{eq:qS}
   \ee
and represent $\cF$ as function of $S$ and $K$.
   \be \label{eq:FS}
   \cF (K, S) = \frac{K}{2}\frac{S^2}{1-S}
   \frac{(1-S)\exp(K S) + 1}{\exp(K S) - 1} + \ln (1-S).
   \ee
   The equations (\ref{eq:uvjeti}) 
are then solved 
in the \cS plane and, by equation (\ref{eq:qS}),
    mapped back into the complex-$q$ plane.
 
The solutions $q^{(m)}_{1,2}$ should comply with both equations  
(\ref{eq:uvjeti}).
  Since the first of equations (\ref{eq:uvjeti}) is $N$-independent, 
   it contains all the possible positions of zeros for arbitrary $N$ 
including the limit $N\to\infty$. The zeros of the system of
particular size $N$ are then determined as intersections of
this curve with solutions of the second equation in (\ref{eq:uvjeti}).
For large $N$ these intersections appear to be equally distributed along the
the curve  $\cF_R   =  0$, so that it can be drawn as the contour 
containing zeros.   (see Fig. 2 as an illustration).
  The curve crosses the positive \req axis
   at a point which is transition point for a given temperature.

Since the locus of zeros has a shape similar to a circle,
we draw the circular contour (long dashed line) in order  to stress
the difference. 
The parameters of this circle are determined by the facts, that its 
center lies on the \req axis (since zeros appear in complex conjugate 
pairs only) and that it intersects the positive \req axis at $q$ solution 
of (\ref{eq:Kexact1}) and the negative \req axis at $q = - |q|$, 
solution of
     \be
      K = 2\,\frac{|q|+1}{|q|+2}\,\ln (|q|+1).   \label{eq:Kexact3}
     \ee   

We were not able to give analytical expressions for all the zeros 
of the model, but the location of zeros closest to the positive real 
$S$ or $q$ axis is a relatively easy task.

Since we limit here to the first-order transitions, 
   the temperature will be fixed through the parameter $Q_0 > 2$
   \be \label{eq:KQ}
    K = 2\,\frac{Q_0-1}{Q_0-2}\,\ln (Q_0-1),
   \ee
In the thermodynamic limit the second minimum of $\cF (S)$
   appears at $S_0 = (Q_0-2)/(Q_0-1)$.
   For large but finite $N$, we will expand free energy (\ref{eq:FS})
   around the minimum, $S = S_0 + s$.
   The small $s \ll S_0$ vanishes as $N\,\to\,\infty$, so we
   expect to have, to leading order in $1/N$,
   \be \label{eq:s1s2}
   s = s_1 + i\,s_2, \hspace{1cm} s_1 = \frac{a}{N^{x_1}},
                     \hspace{0.5cm} s_2 = \frac{b}{N^{x_2}},
   \ee
   for real constants $a, b, x_1, x_2$.
   The expansion of (\ref{eq:FS}) and the constraint (\ref{eq:uvjeti})
   give the equation
   \bea
   \cF & = & (Q_0 - 1)\,\left[
    1 - \ln\,(Q_0 - 1)\,(A_1 - B_1)
         \right] \, s + \cO(s^2)
   = 0 \pm\,i\,\frac{2\,m + 1}{N}\,\pi, \nonumber \\
   A_1 & = &  \left(\frac{K - Q_0 + 1}{Q_0} +
     \frac{2}{Q_0 - 2} \right), \;
   B_1 = \frac{Q_0 - 1}{Q_0\,(Q_0 - 2)}\,K - 1.
   \eea
   The closest zero has the smallest imaginary
   part, so its index is $m = 0$.
   By inserting (\ref{eq:s1s2}) in the above equation,
   one reads
   \be \label{eq:konv1}
   a = 0, \hspace{1cm}
   b = \frac{\pi}{(Q_0 - 1) - \ln\,(Q_0 - 1)\,(A_1 - B_1)},
   \hspace{1cm} x_2 = 1,
   \ee
   while $x_1$ is not defined.
Mapping back to the complex-$q$ plane by using the equation (\ref{eq:qS})
gives
   \bea \label{eq:fp1qN}
   q_1^{(0)}(N) & = & Q_0 + \frac{(Q_0 - 1)^2}{Q_0 -2}\,
   (K - Q_0)\,\frac{a}{N^{x_1}} \nonumber \\
   & + & \frac{(Q_0 - 1)^2}{Q_0 -2}\,\left\{
   \frac{K}{2}\,[K - 2\,(Q_0 - 1)] + \frac{Q_0 - 1}{Q_0 - 2}\,(Q_0 - K)
   \right\}\,\left(
   \frac{a^2}{N^{2\,x_1}} - \frac{b^2}{N^{2\,x_2}}
   \right) \nonumber \\
   & + & \cO\left(N^{-3\,x}\right), \\
   q_2^{(0)}(N) & = &  \frac{(Q_0 - 1)^2}{Q_0 -2}\,
   (K - Q_0)\,\frac{b}{N^{x_2}} \nonumber \\
   & + & \frac{(Q_0 - 1)^2}{Q_0 -2}\,\left\{
   \frac{K}{2}\,[K - 2\,(Q_0 - 1)] + \frac{Q_0 - 1}{Q_0 - 2}\,(Q_0 - K)
   \right\}\,\frac{2\,a\,b}{N^{x_1 + x_2}}  \nonumber \\
   & + & \cO\left(N^{-3\,x}\right). \nonumber
   \eea
   Finally, by inserting values for $a, b, x_2$ from (\ref{eq:konv1}),
   one obtains the large-$N$ dependence of the 
distance of the closest zero from the real axis in the case of the 
first-order transitions.
   \bea \label{eq:q12N}
   q_1^{(0)}(N) & = & Q_0 - \frac{(Q_0 - 1)^2}{Q_0 -2}\,\left\{
   \frac{K}{2}\,[K - 2\,(Q_0 - 1)] + \frac{Q_0 - 1}{Q_0 - 2}\,(Q_0 - K)
   \right\}\,\frac{b^2}{N^2} \nonumber \\
   & + &  \cO\left(N^{-3\,x}\right), \nonumber \\
   q_2^{(0)}(N) & = & \frac{(Q_0 - 1)^2}{Q_0 -2}\,
   (K - Q_0)\,\frac{b}{N} + \cO\left(N^{-3\,x}\right),
   \eea
   The above expressions can be checked numerically. 
For illustration,  the numerical data of the closest zeros, 
   for the temperature
critical for $q=8$
   are presented in Table 2 
for different  sizes $N$.

    \begin{table}[htb]  \label{tb:Nq8}
    \caption{ The location of the closest zeros for the temperature 
given by the relation (\ref{eq:KQ}) for $Q_0 = 8$.  $N$ denotes the 
number of  the system considered.}
    \begin{center} \begin{tabular}{ccc}
    &  & \\
    \hline
     $N$ & $q_1^{(m = 0)}(N)$ & $q_2^{(m = 0)}(N)$  \\
   5000&7.99999543669504&$ -8.679091351481774\,\cdot\,10^{-3}$ \\
   6000&7.99999683103614&$ -7.232577480179549\,\cdot\,10^{-3}$ \\
   7000&7.99999767178073&$ -6.199352825621011\,\cdot\,10^{-3}$ \\
   8000&7.99999821745666&$ -5.424434119811201\,\cdot\,10^{-3}$ \\
   9000&7.99999859157044&$ -4.821719459791061\,\cdot\,10^{-3}$ \\
  10000&7.99999885917194&$ -4.339547669719317\,\cdot\,10^{-3}$ \\
  11000&7.99999905716679&$ -3.945043440975050\,\cdot\,10^{-3}$ \\
  12000&7.99999920775815&$ -3.616289894012384\,\cdot\,10^{-3}$ \\
  13000&7.99999932495365&$ -3.338113800831173\,\cdot\,10^{-3}$ \\
  14000&7.99999941794470&$ -3.099677139486969\,\cdot\,10^{-3}$ \\
  15000&7.99999949296515&$ -2.893032025972416\,\cdot\,10^{-3}$ \\
  \hline \end{tabular} \end{center}
   \end{table}

The data $q_{1, 2}^{(0)}(N)$ from Table 2 and analogous series of data for 
different temperatures corresponding to $q=3$ and $q=2.1$ were extrapolated 
to the thermodynamic limit by using the BST algorithm. 

The extrapolation  parameter $\omega$ was fixed by the constraint 
$q_2^{(0)}(N)\,\to\,0$.
After we have made the extrapolations of $q_{1, 2}^{(0)}(N)$
and found the values of $q_1(\infty)$ and $\omega$, we have performed 
alternative calculation of convergence exponents of  $q_{1, 2}^{(0)}(N)$.
By assuming the convergence of the $q_{1, 2}^{(0)}(N)$ to be of the form  
   \be \label{eq:q12alfa}
    q_1^{(0)}(N)  - 
    q_1(\infty) = C_1\,\cdot\,N^{-\alpha_1 } + \cdots,
    \hspace{0.9cm}
    q_2^{(0)}(N) = C_2\,\cdot\,N^{-\alpha_2 } + \cdots,
   \ee
the convergence exponents $\alpha_{1, 2}(N)$ for real and imaginary part 
of the closest zeros, can be expressed as
   \bea \label{eq:alfa12}
   \alpha_1(N) & = &  - \frac{\ln\left\{\left[q_1^{(0)}(N) - q_1(\infty)\right]/
   \left[q_1^{(0)}(N + 1) - q_1(\infty)\right]
   \right\}}{\ln[N/(N + 1)]},  \nonumber \\
   \alpha_2(N) & = & - \frac{\ln\left[
   q_2^{(0)}(N) / q_2^{(0)}(N + 1) \right] }{\ln[N/(N + 1)]} .
   \eea
   The above data for $\alpha_{1, 2}(N)$ are extrapolated
again by the BST algorithm.
   The complete set of extrapolated results is summarised 
in the Table 3. 

   \begin{table}[hbt] \label{tb:bst1ord}  
   \caption{ The results of the BST extrapolations performed on the
    closest zeros, $q_{1, 2}(\infty)$, followed by parameter
    $\omega$ and convergence exponents $\alpha_{1, 2}$,
    for set of temperatures determined by $Q_0$ and (\ref{eq:KQ}).
    The predicted convergence exponents $x_{1, 2}$ as given
    by (\ref{eq:q12N}) are in the
    last two columns. }
    \begin{center}
    \begin{tabular}{ cccccccc }
    & & & & & & & \\
    \hline
   $Q_0$ & $q_1(\infty)$ & $q_2(\infty)$ & $\omega$ &
   $\alpha_1$ & $\alpha_2$ & $x_1$ & $x_2$  \\
    8 & 8.0000000000001 & $3\,\cdot\,10^{-13}$ & 1.0000 &
    1.9999988 & 1.000000002 & 2 & 1 \\
    3 & 3.000000000005 & $4\,\cdot\,10^{-12}$ & 1.00000 &
    2.00003 & 0.99999997 & 2 & 1 \\
    2.1 & 2.1002 & $9\,\cdot\,10^{-6}$ & 0.98 & 1.85 & 0.95 & 2 & 1 \\
    \hline
    & & & & & & & \\
   \end{tabular}
   \end{center}
   \end{table}

  We observe a full agreement between the analytically given values for 
$q$ and convergence exponents on one side, and the numerically
  calculated and extrapolated values for $q$ and exponents on the other side.
Although the scaling in the case of the first-order phase transition 
is trivial and linearly proportional to system size, the results can be used
to check the convergence of numerical results. 
For $q=3$ and $q=8$ the convergence exponent has improved compared to 
the earlier numerical results given in Table 1, confirming that larger 
error in this case comes from the small size of considered systems and 
that scaling is indeed linear with size as expected for the first-order 
phase transition.
   The loss of precision in extrapolations persists only for the
case corresponding to $q=2.1$ and should be attributed to the 
crossover effects due to vicinity of  the border between the first- 
and second-order transitions.

     \vspace{1cm}

     For the partition function zeros lying on the curve in the appropriate
     complex plane, the normalised density of zeros $g$, is defined as
      \cite{LY52}
     \be
     \label{eq:defg}
     g = \frac{1}{N}\,\frac{d n}{d l}, \hspace{1cm} \int\,g\,dl = 1,
     \ee
     where $d n$ is the number of zeros inside the arc of the curve of
     the length $d l$. In the case of the field- or temperature-driven
     second-order transitions, the density of zeros vanishes 
at the transition with an exponent connected to
     critical exponents \cite{CK97}. On the contrary, at the first-order
     transition point, the density of zeros remains constant.
     To the end of this paragraph, we will examine the behaviour of 
density of complex-$q$ zeros in the vicinity of first-order transition point.
     To this purpose, we notice that the large-$N$ expansion
     of $q_{1, 2}(N)$ for the closest zeros (those with $m =0$)
presented earlier in text, 
 may be directly generalised to other values of $m$
as long as  the condition $(2 m + 1) \ll N$ is fulfilled. 
This leads to
   \bea
    q_1^{(m)}(N) - q_1(\infty) & = &
     C_1\,\cdot\,\left(
    \frac{2 m + 1}{N}\right)^{\alpha_1}  + \cdots,
     \nonumber \\
    q_2^{(m)}(N) & = & C_2\,\cdot\,\left(
    \frac{2 m + 1}{N}\right)^{\alpha_2} + \cdots,
   \eea
     with the constants $q_1(\infty), C_1$ and $C_2$ given by
     the relations (\ref{eq:q12N}), where $1/N$ is changed by
     $(2 m + 1)/N$.

     To calculate numerically density of zeros (\ref{eq:defg}), 
     we choose two neighbouring zeros, with $d l$ denoting 
     geometrical distance between them, \\
     $ d l = \left\{ \left[q_1^{(m + 1)} - q_1^{(m)}\right]^2 +
     \left[q_2^{(m + 1)} - q_2^{(m)}\right]^2\right\}^{1/2}$.
     In that case density of zeros near positive real-$q$ axis becomes
     \bea
     g(m) = & \frac{1}{N} & \,\left\{
     C_1^2\,\left(\frac{2 m + 1}{N}\right)^{2\,\alpha_1}\,\left[\left(1 +
     \frac{2}{2 m + 1}\right)^{\alpha_1} - 1\right]^2
     \right. \nonumber \\
     & + & \left.
     C_2^2\,\left(\frac{2 m + 1}{N}\right)^{2\,\alpha_2}\,\left[\left(1 +
     \frac{2}{2 m + 1}\right)^{\alpha_2} - 1\right]^2
     \right\}^{-1/2}.
     \eea
     The edge $(m = 0)$ density of zeros $g_0$, tends to constant value 
     in the thermodynamic limit.   
     \be \label{eq:gN1}
     g_0 = \frac{1}{2\,C_2}\,\left[
     1 - 8\,\frac{C_1^2}{C_2^2}\, \frac{1}{N^2} + \cdots
     \right] \;\to\; \frac{1}{2\,C_2},
     \ee

For example, for $Q_0 = 8$ and $K(Q_0 = 8) = 4.540457014462$ 
this constant is
     \be \label{eq:constg}
     \frac{1}{2\,C_2} = \left[ 2\,\frac{(Q_0 - 1)^2}{Q_0 -2}\,
     (K - Q_0)\,b \right]^{-1}
      =  0.011521939 \cdots   ,
     \ee
For illustration, we check this result numerically by calculating    
  for the values of $Q_0$ and $K$
     from above, and $N = 10\,000$,  the
     twenty zeros closest to the positive \req axis and corresponding densities.
They are presented in Table 4. 

   \renewcommand              
   \baselinestretch 1              
   \baselineskip 15pt              %
   \begin{table} \label{tb:gq8}
   \caption{ The twenty zeros closest to the positive \req axis
   and the corresponding densities are
   shown at temperature determined from the relation
    (\ref{eq:KQ}) and $Q_0 = 8$ and for the $ N = 10\,000$ spins.}
   \begin{center} \begin{tabular}{cccc}
      & & & \\
   \hline 
  $m$ & $q_1^{(m)}$ & $q_2^{(m)}$ & $g(m)$ \\
  0&7.99999885917192&-0.004339547669719&0.0115219529910602 \\
  1&7.99998973259138&-0.013018627057442&0.0115219974088187 \\
  2&7.99997147966529&-0.021697658591626&0.0115220714335465 \\
  3&7.99994410086352&-0.030376610375500&0.0115221750581617 \\
  4&7.99990759689083&-0.039055450520507&0.0115223082726076 \\
  5&7.99986196868649&-0.047734147149695&0.0115224710640261 \\
  6&7.99980721742419&-0.056412668400979&0.0115226634165474 \\
  7&7.99974334451167&-0.065090982430549&0.0115228853117394 \\
  8&7.99967035159032&-0.073769057415945&0.0115231367280630 \\
  9&7.99958824053479&-0.082446861559540&0.0115234176412815 \\
 10&7.99949701345249&-0.091124363091702&0.0115237280243068 \\
 11&7.99939667268302&-0.099801530274076&0.0115240678473335 \\
 12&7.99928722079755&-0.108478331402754&0.0115244370775645 \\
 13&7.99916866059815&-0.117154734811648&0.0115248356796385 \\
 14&7.99904099511710&-0.125830708875537&0.0115252636152771 \\
 15&7.99890422761603&-0.134506222013369&0.0115257208435568 \\
 16&7.99875836158505&-0.143181242691353&0.0115262073207999 \\
 17&7.99860340074195&-0.151855739426127&0.0115267230005962 \\
 18&7.99843934903115&-0.160529680787899&0.0115272678338162 \\
 19&7.99826621062264&-0.169203035403569&0.0115278417687404  \\ 
   \hline 
      & & & \\
   \end{tabular} \end{center} \end{table}
   \renewcommand              
   \baselinestretch 1              
   \baselineskip 15pt              %

     The limiting value of the density at the transition point, is extracted
      by the BST extrapolation procedure (\ref{eq:seq}) simply
     by changing variable $1/N$ into $(2\,m + 1)/N$.
     \be \label{eq:g0}
     g_m = g_0 +{\tilde g}\,\left(\frac{2\,m + 1}{N}\right)^{\omega} + \cdots.
     \ee
     The extrapolations give $g_0 = 1.152194\,\cdot\,10^{-2}$ for a 
     wide range of $0.3 < \omega < 3.0$, which reproduces the analytically 
obtained value (\ref{eq:constg}) in six significant digits. 

 We can conclude that the density of the partition function zeros in the
 complex-$q$
  plane behave in similar way as the density of zeros in the
  complex-temperature \cite{F65} or complex-field \cite{CK97} planes:
  in thermodynamic limit
  it has constant value at the first-order transition point.

\section{The power-law  decaying interactions} \label{sec-lr}  

The second case that we have considered is the ferromagnetic Potts model 
in one dimension with power-law decaying interactions. When taking the 
periodic boundary conditions, it is described by the reduced Hamiltonian
  \be
   \cH_{PL} = - \frac{H_{PL}}{k_B\,T} = 
   K \sum_{i=1}^{N-1}\;\sum_{j=1}^{N-i}\; \delta\,(s_i, s_{i+j}) 
   \left[\frac{1}{j^{1+\sigma}}+\frac{1}{(N-j)^{1+\sigma}}\right]. 
  \ee
Although in one dimension, the model is nontrivial and has a phase 
transition at nonzero temperature for all $q$ when 
 $0 < \sigma \leq 1$ \cite{D69,ACCN88,GU93}. 
This transition is of the mean-field type for low enough values of 
$\sigma$,  $\sigma< \sigma_c(q)$, and is a
non-trivial second-order phase transition for $\sigma > \sigma_c(q)$.
In the classical regime $\sigma < \sigma_c(q)$, the transition is of the 
first order when $q > 2$ \cite{UG97,GU98,BDD99}.
The exact position of the line separating the two regimes, $\sigma_c(q)$, 
 is a difficult and still open question \cite{GU98,BDD99,UG00}.

 In  the graph representation, the difference from the mean-field case is that 
the contribution of active links depends on distance between the spins 
they connect.
The contribution of active bonds may then have $N/2$ (for $N$ even) 
 or $(N - 1)/2$ (for $N$ odd) distinct values
given by $v_j = \exp[K/j^{1+\sigma}+K/(N-j)^{1+\sigma}] - 1$.
If we denote by  $b_j(\cG)$ the number of active links of the length 
$j$ or $N - j$ in a graph $\cG$, the partition function becomes  
   \be \label{eq:zlr}   
   Z_N = \sum_{{\rm all}\;\cG} \, v_1^{b_1(\cG)}\,v_2^{b_2(\cG)}\,\cdots
   \,q^{n(\cG)} 
   \ee
    or, in form of polynomial in $q$, 
   \be \label{eq:Npollr}   
   Z_N = \sum_{n = 1}^N\,a_n(v_j, N)\;q^n.
   \ee
   The real and positive coefficients $a_n$ depend on temperature and  $N$, 
but also on the parameter of range $\sigma$. 
This makes the calculations more demanding for the computer memory.

 We have performed exact numerical calculations on chains with 
$N \leq 9$ spins for two sets of parameters,  
 ($\sigma = 0.3$, $K = 0.576$) and ($\sigma = 0.8$ $K = 0.8230$). 
As in the previous case, the chosen temperatures are the critical
temperatures known for certain values of $q$. They correspond to 
$q = 5$ and  $q = 2$ respectively.
Remark, however, that in the present case the critical temperatures
are not known exactly. The values used here were obtained by the finite-range 
scaling approach \cite{GU93}.
The first and second set correspond respectively to the  to the first- 
and second-order phase transition.

The numerical results for the  loci of complex-$q$ zeros for sizes
 $N = 6, \cdots, 9$ 
are presented in Figures 3 and 4.
  In both examples are obtained the zeros lying on the arc-shaped curves,
while the positive real axis is free of zeros. The zeros closest to it
approach to it with increasing $N$.

As in the mean-field case, we have examined the convergence of the
closest zeros
to the thermodynamic limit  by the BST algorithm, assuming the power-law form.
The analysis included the data obtained for chains with $N = 5$ to $9$. 
The parameter $\omega$ was determined, as earlier, by requiring that
 $q_2 \,\to\,0$ in the thermodynamic limit. 
The results are presented in Table 5. 

   \renewcommand              
   \baselinestretch 1              
   \baselineskip 15pt              %
   \begin{table}[hbt]  \label{tb:bstLR}
   \caption{ The loci of closest zeros and its BST extrapolation at
    $(K = 0.576, \sigma = 0.3)$ and $(K = 0.823, \sigma = 0.8)$} 
   \begin{center} \begin{tabular}{cccc} \hline
      & $ K = 0.576$ & $\sigma = 0.3$ &  \\ \hline  
  $N$ & $q_1^{(0)}(N)$ &  $q_2^{(0)}(N)$ & $\omega$  \\
    5 & -0.447413829093298& 2.78806303559461 & \\
    6 & 0.0734103695976816& 2.88316232503579 & \\
    7 & 0.486108948338677& 2.88942370075814 & \\
    8 & 0.818954244961081& 2.85269647435491 & \\   
    9 & 1.09269817849075& 2.79489503015230 & \\ 
 $\infty$ & 4.2        & $3\,\cdot\,10^{-3}$  & 0.89 \\ \hline 
    & $ K = 0.823$ & $\sigma = 0.8$  \\ \hline
  $N$ & $q_1^{(0)}(N)$ &  $q_2^{(0)}(N)$ & $\omega$ \\  
    5 &-0.418572384698087 & 3.69577294340190 & \\
    6 & 0.179124021423254 & 3.55212427658784 & \\    
    7 & 0.581821465484680 & 3.37723067427777 & \\   
    8 & 0.866098195024489 & 3.20669717997192 & \\
    9 & 1.074705295884560 & 3.05042285906340 & \\  
 $\infty$ & 2.01       & $3\,\cdot\,10^{-3}$  & 0.41 \\  
  \hline \end{tabular} \end{center} \end{table}
   \renewcommand              
   \baselinestretch 1              
   \baselineskip 15pt              %
 
In the case of second order phase transition the expected 
crossing point with the real axis is obtained with precision better than 1\%.
The leading convergence exponent given by $\omega$, 
is equal to 0.41, 
which is close to the value for the temperature critical exponent 
$1/\nu =0.48$ obtained earlier by finite-range scaling
for $q=2$ and $\sigma=0.8$ \cite{GU93}. 

In the case of the first-order transition the fit is more difficult.
The discrepancy between the obtained crossing point and the
value expected for the considered temperature 
is larger, about 20\%, while the convergence exponent, which 
in context of finite-size scaling should be linear for a first-order 
phase transition, differs by about 11\%. 
Larger error in determination
of the value of $q$, may be attributed to both lower precision of
the data for critical temperature obtained by finite-range scaling
 \cite{GU93,GU98} 
when the values of $\sigma$ are low and to  the proximity of
the second-order transition regime in the $(q,\sigma)$ plane, 
which might induce crossover effects for small sizes.

 Since in these preliminary calculations we dispose with small number 
of data, we have not calculated the density of zeros in the case of 
  power-law model.

  \section{Conclusion} \label{sec-zklj}  

  We have studied the \cq zeros of the partition function of the ferromagnetic
  Potts model with long-range interactions in context of the first- and
  second-order phase transitions in this model.
  Two cases are presented: the mean-field limit and a few examples
  of the case with power-law decaying interactions.

  We conclude that the positions of the zeros may be analysed in a 
  similar way as the zeros in the plane of complex field or 
  temperature and that they can give useful informations on the 
  phase transition in ferromagnetic Potts models.   

Numerical results for relatively small systems show the similar picture 
as in the case of zeros in the complex field or temperature plane. 
In all the cases considered the zeros lie on smooth curves
that intersect the real positive axis at value for which the given 
temperature is the transition temperature.
For finite systems the positive real axis is free of zeros, but they 
approach to it with increasing size.

 The distance of the closest zero from the real axis is found to scale
 with size as the temperature critical exponent $1/\nu$.
 In the mean-field case, this result could be obtained  with a good precision
 for series of different temperatures corresponding to different transition
 regimes. In the case of power-law interactions, the preliminary results
 obtained with less precision suggest the same behaviour.
 This result is similar to behaviour in the complex temperature plane,
 and it supports recent findings by Kim and Creswick \cite{KC01}
 in the short-range Potts model in $2d$.  

In the end, let us mention some technical advantages of the analysis 
by complex-$q$ zeros compared to the complex-temperature zeros
in the model with long-range interactions.
  The partition function of the mean-field model consisting of $N$
   spins can
be written as a $N$-th order polynomial in $q$, or roughly 
$N^2$-th order polynomial in temperature variable. Consequently, 
this allows better numerical precision of calculations in the complex-$q$
plane then those in the complex-temperature plane.
  For a model with power-law decaying interactions the possibilities 
in complex temperature plane are even more restricted. 
Model with $N$ spins, has $N/2$ (for $N$ even) 
 different parameters involving 
temperature, so that the partition function cannot be defined as a 
polynomial in some temperature variable at all.

 Further investigations of the complex-$q$ zeros should be done in
 future in at least two directions.
 We plan to perform a more extended numerical calculations of the
 complex-$q$ zeros in  the case of power-law decaying interactions.
 This should allow us to obtain the convergence exponent for the closest 
 zero with more  precision. 
 In addition, it would be interesting to examine a still
 open question of the borderline $q_{c}(\sigma)$ dividing
 the first- from the second-order transition regimes within the
 formalism of complex-$q$ zeros.
 Also, on the basis of the results obtained in this paper, a possible
 connection between the zeros in \cq and complex-temperature
 plane deserves further investigation.

  \newpage

  \centerline{\Large Figure caption}

  \vspace{1cm}

  {\bf Fig. 1}: The loci of complex-$q$ zeros of the partition function of
              the mean-field Potts
              model calculated at inverse temperatures
              (a) $K = K_c(q = 0.5) = 0.5$, (b) $K = K_c(q = 2) = 2$ and 
              (c) $K = K_t(q = 8.0) = 4.540457$ for 
              $N = 25$ (open diamonds), $N = 30$ (filled triangles) 
              and $N = 35$ (filled circles) spins.

  {\bf Fig. 2}: The locus of complex-$q$ zeros of the partition function of
              the mean-field Potts
              model calculated at inverse temperature
              $K = K_{MF}(q = 8) = 4.540457$.     
              The full line denotes loci of zeros in the limit $N\to\infty$, 
obtained by the approach by the saddle point approach
as a solution of the equation $\cF_R = 0$.
   Long dashed line is a circle (described in text) drawn as a guide to the eye.

  {\bf Fig. 3}: The complex-$q$ zeros of the partition function of
                the $1 d$ 
Potts model with power-law interactions at the $FRS$ estimate of first-order
              transition temperature
              $K = K_t(q = 5, \sigma = 0.3) = 0.576$,
              for 
              $N = 6$ (open diamonds), $N = 7$ (filled triangles),
              $N = 8$ (open squares) and $N = 9$ (filled circles) spins.

  {\bf Fig. 4}: The complex-$q$ zeros of the partition function of the $1 d$ 
        Potts model with power-law interactions at the $FRS$ 
        estimate of second-order transition temperature  
              $K = K_c(q = 2, \sigma = 0.8) = 0.8230$,
              for 
              $N = 6$ (open diamonds), $N = 7$ (filled triangles),
              $N = 8$ (open squares) and $N = 9$ (filled circles) 
              spins.

\vfill
\eject

   \centerline{\includegraphics{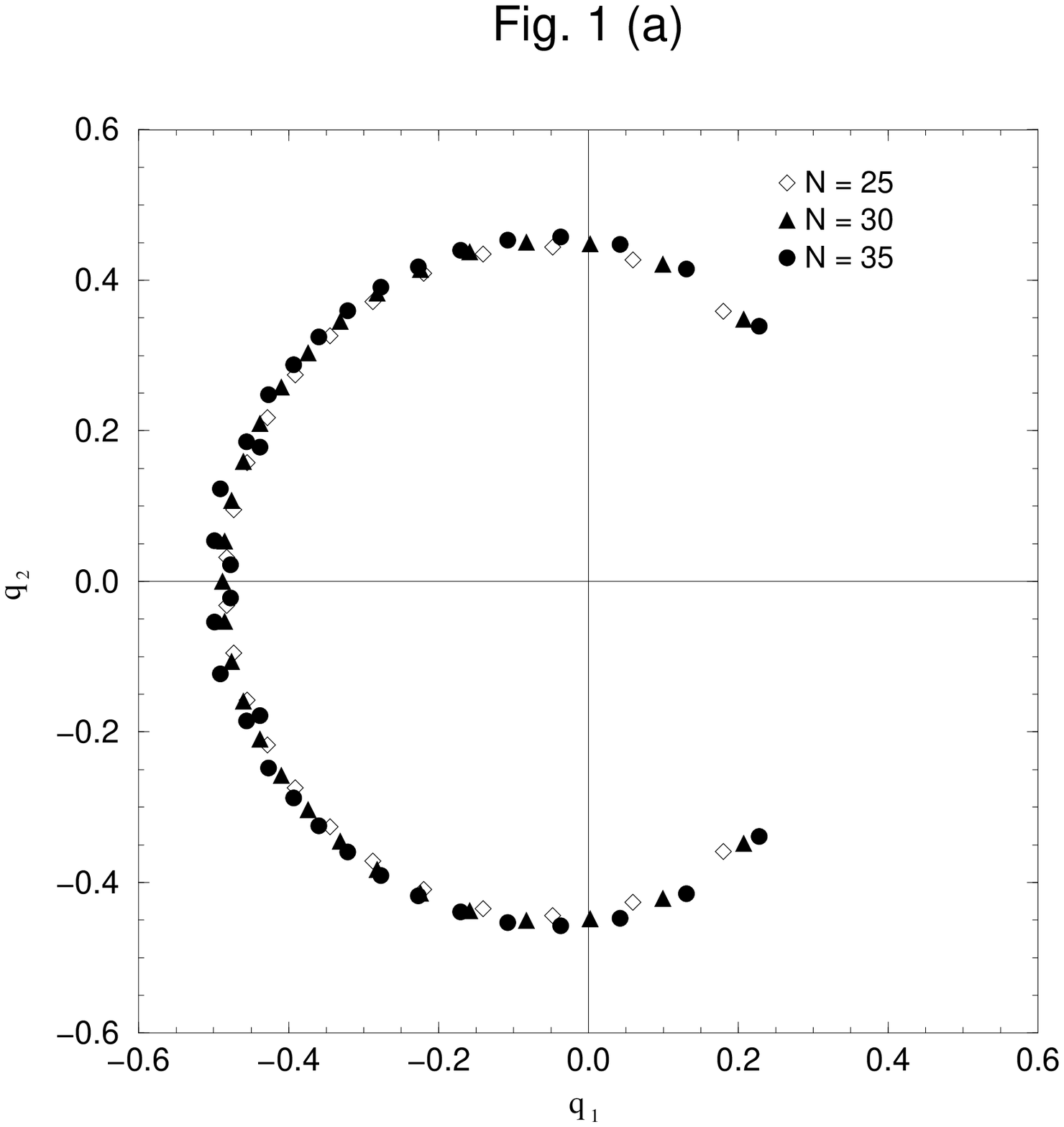}}

   \centerline{\includegraphics{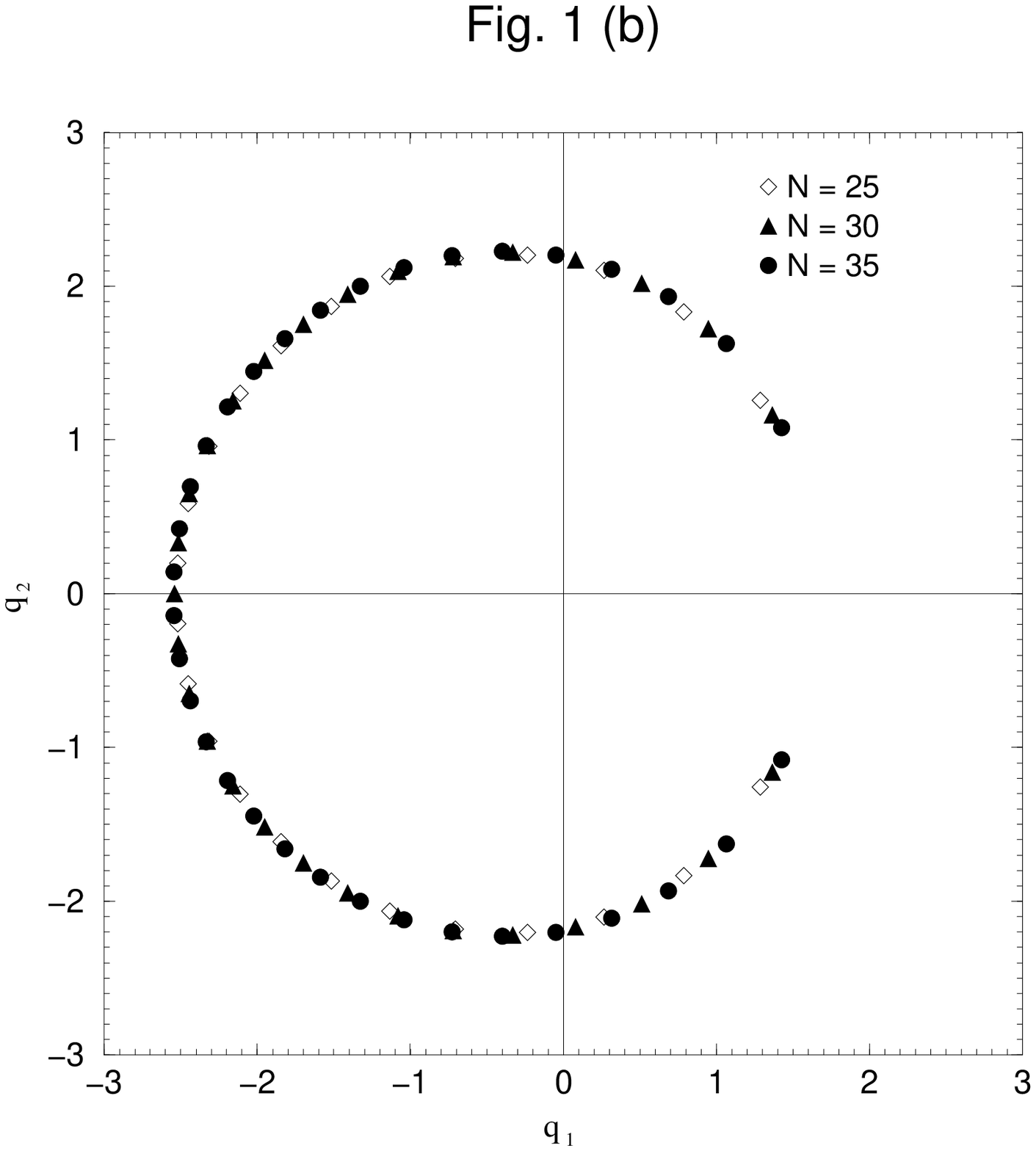}}

   \centerline{\includegraphics{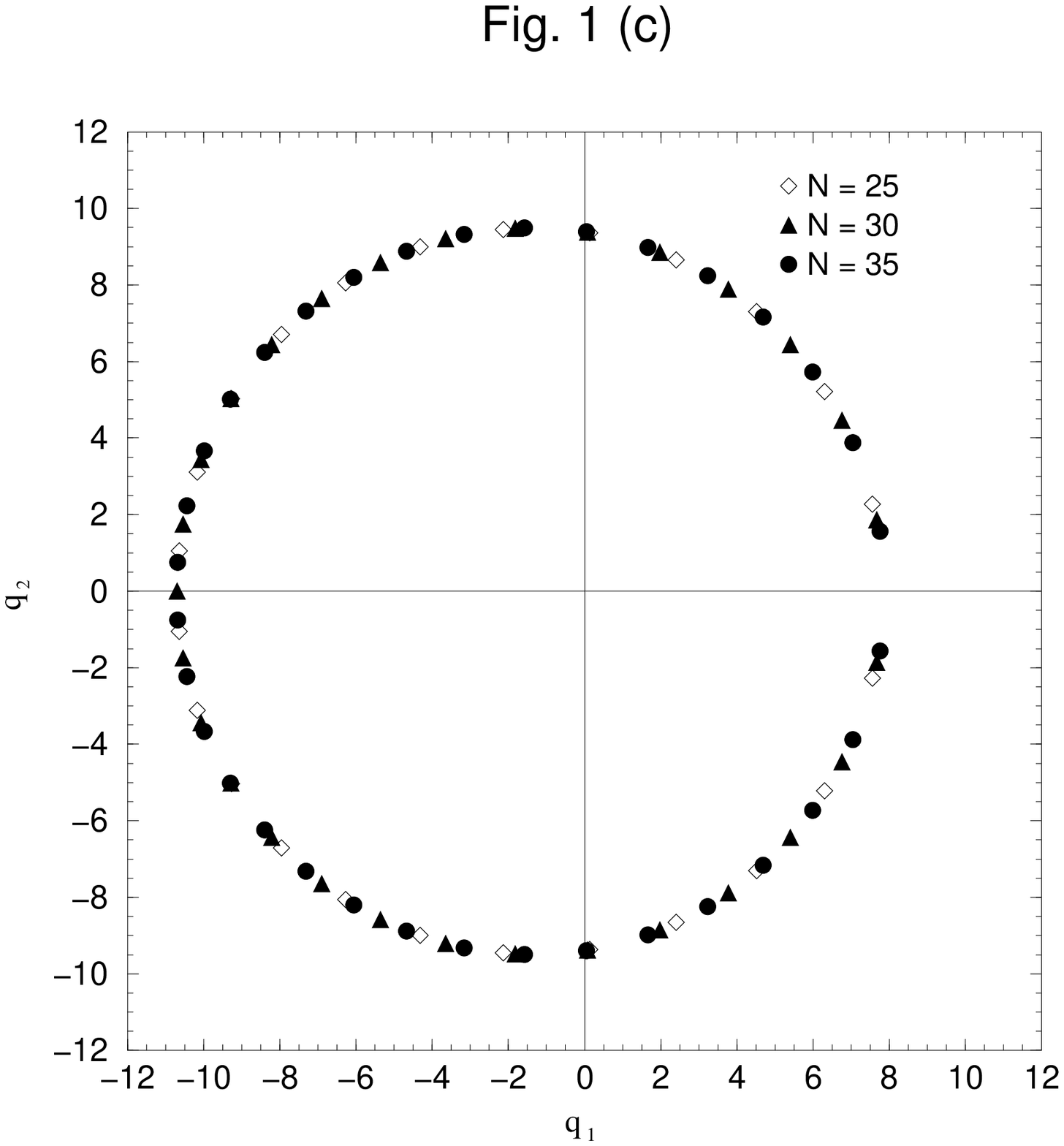}}

   \centerline{\includegraphics{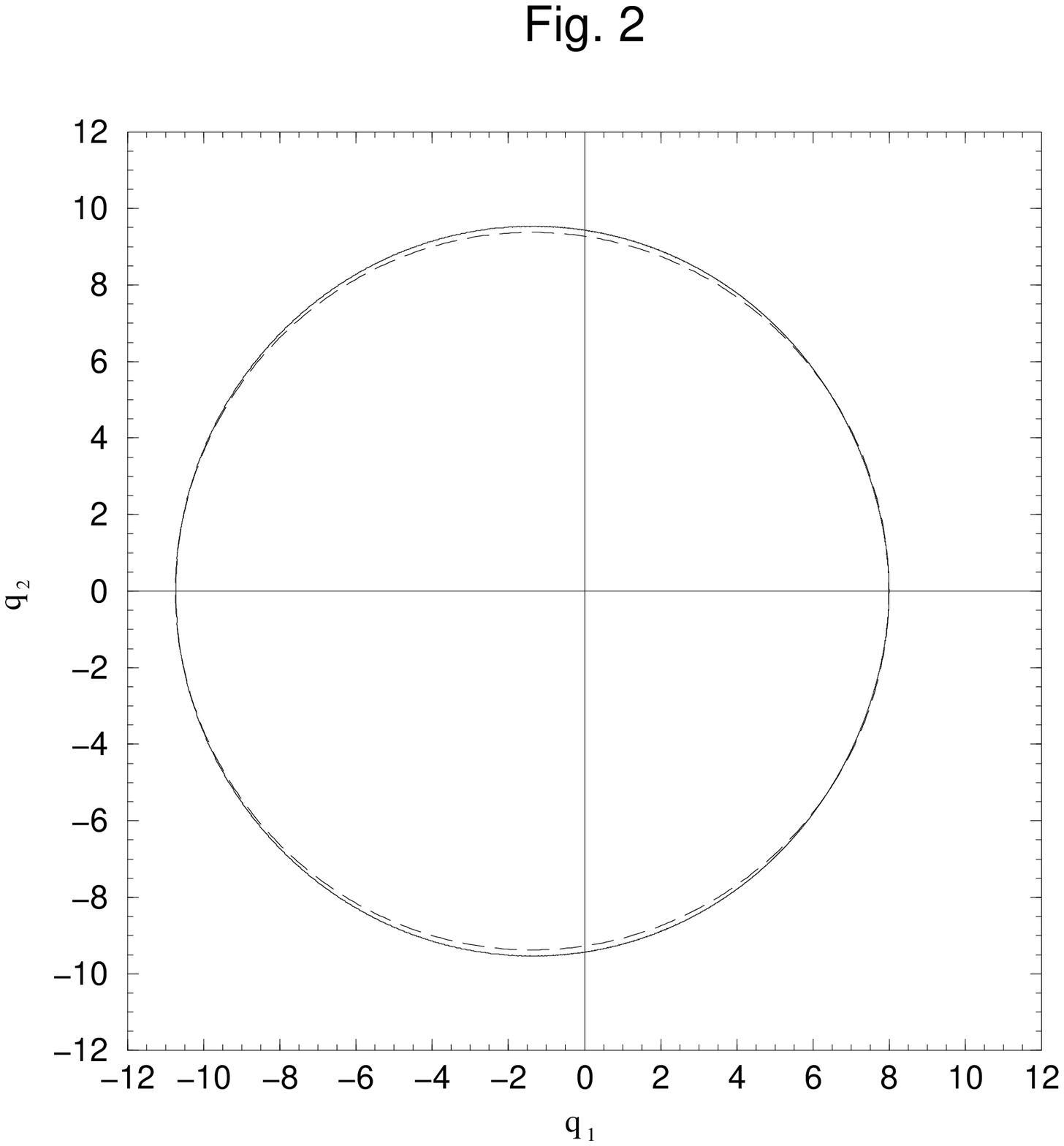}}

   \centerline{\includegraphics{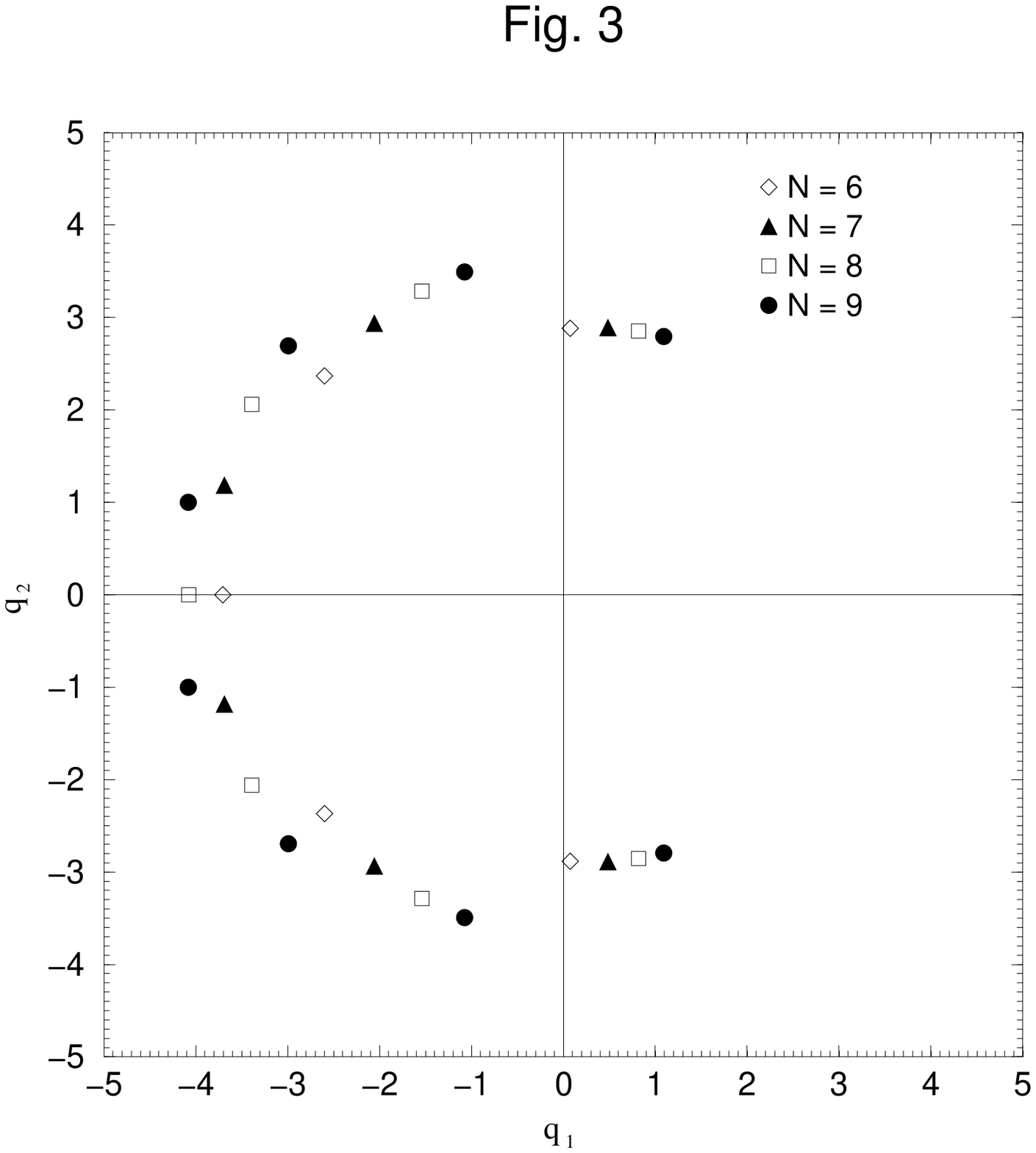}}

   \centerline{\includegraphics{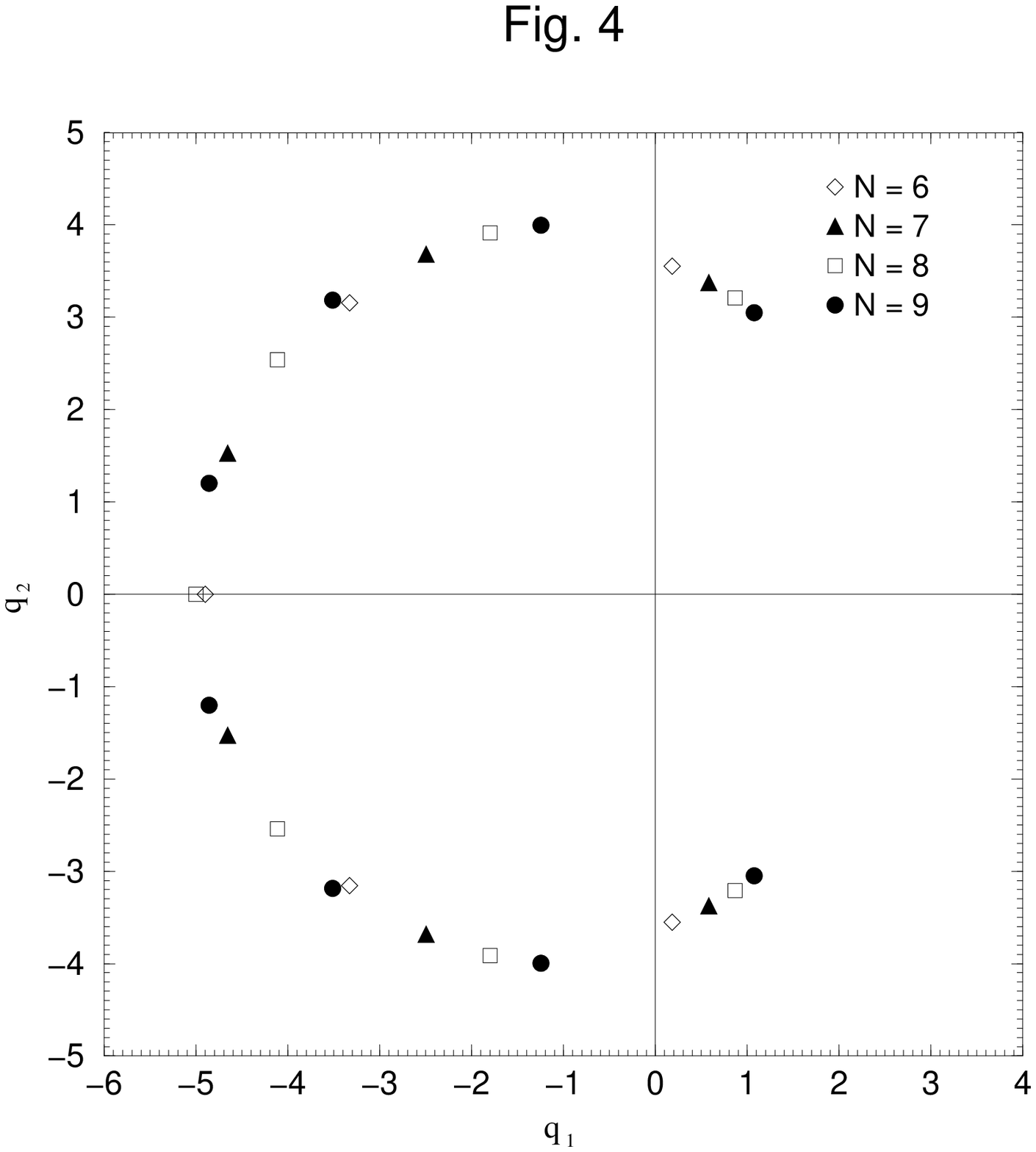}}

    \end{document}